\documentclass[twocolumn,amsmath,nofootinbib]{revtex4}

\usepackage[dvips]{color}
\usepackage{latexsym} 
\usepackage{amssymb}  
\usepackage{graphicx}
\newcommand{\hide}[1]{}

\begin{document}

\title{On Math, Matter and Mind}

\author{Piet Hut$^1$, Mark Alford$^2$ \& Max Tegmark $^3$}

\affiliation{$^1$Institute for Advanced Study, Princeton, NJ 08540}
\affiliation{$^2$Dept.~of Physics, Campus Box 1105, Washington University, St Louis MO 63130}
\affiliation{$^3$Dept.~of Physics \& MIT Kavli Inst.~for Astrophysics and Space Research, MIT, Cambridge, MA 02139}
\date{Submitted to Foundations of Physics October 20, 2005; accepted December 21}  

\begin{abstract}
We discuss the nature of reality in the ontological context of
Penrose's math-matter-mind triangle. The triangle suggests the
circularity of the widespread view that math arises from the mind,
the mind arises out of matter, and that matter can be explained
in terms of math. 
Non-physicists should be wary of any claim that
modern physics leads us to any particular resolution of this circularity,
since even the sample of three theoretical physicists writing this paper 
hold three divergent views.
Some physicists believe that current physics has
already found the basic framework for a complete description of
reality, and only has to fill in the details. Others suspect that no
single framework, from physics or other sources, will ever capture
reality.  Yet others guess that reality might be approached
arbitrarily closely by some form of future physics, but probably based
on completely different frameworks. We will designate these three
approaches as the fundamentalist, secular and mystic views of the
world, as seen by practicing physicists.  We present and contrast each
of these views, which arguably form broad categories capturing
most if not all interpretations of physics.
We argue that this diversity in the physics community 
is more useful than an ontological monoculture, since it motivates physicists to tackle 
unsolved problems with a wide variety of approaches.
\end{abstract}

\maketitle


\section{Introduction: the Role of Metaphor}
\label{sec:intro}

Although physicists agree on the formalism of their theories and the
methodology of their experiments, they often disagree about the
question of what it all means.
A few hundred years ago, this was only to be
expected, since at its inception, physics covered very limited aspects of
the world. However, now physics can arguably lay proper
claim to providing an effective description of all material processes,
if not in practice then at least in principle.  And with no obstacles
visible to a full understanding of the dance of matter and energy in
space and time, what aspect of the world would not be amenable to an
analysis by physics, again at least in principle?

The obvious objection would be to point out that notions such as
meaning or beauty or responsibility have been explicitly filtered out
of physics, as part of its methodology.  To believe that you could
remove some of the most important aspects of human experience, and
then hope to fully reconstruct them through the mathematical formalism
of physics strikes many as absurd.  This view, that physics can only
cover limited aspects of our experience, we will call the secular
view.  The opposite view that holds that a straightforward application
and further exploration of the current framework of physics will
eventually cover and explain all of reality we will call the
fundamentalist view.

There is a reason for choosing religious terms for our metaphors.
First of all, they indicate a mind set, a psychological attitude that
corresponds quite closely to the ones we will discuss among physicists.
Secondly, they are universal, in that the tension between the more
fundamentalist and the more secular views of religion occur everywhere,
in Christianity, Judaism, Islam, Hinduism, Buddhism, and so on.

\begin{figure}
\begin{center}
\includegraphics[width=3in]{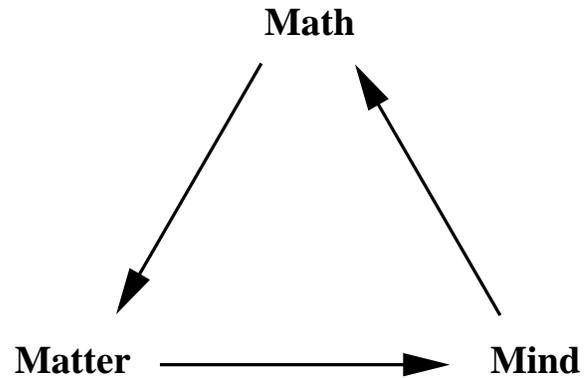}
\end{center}
\caption{Our starting diagram, based on a similar one
by Penrose \cite{Penrose}}
\label{fig:mmm}
\end{figure}

There is one other type of viewpoint that is widespread across
religions, and that is the mystic view of the world.  Many religions
have produced mystics
who use precise methods and descriptions in
order to find deeper forms of truth that will eventually transcend
both methods and descriptions, leading to insight that is utterly
down-to-earth but that cannot be captured by the net of description.

Among physicists, the mystic view holds
that a future form of physics may come arbitrarily close to a genuine
understanding of reality, both from a pragmatic and an ``insight'' point
of view.  However, to what extent this insight can be captured in any
way through presently familiar forms of mathematical physics or any
other future form of physics relying on description, is not clear and
is left as an open question.  This may sound presumptuous, but there
are good historical reasons to expect fundamental changes in what is
considered legitimate physics.  For example, the discovery of quantum
mechanics toppled some of the notions held most near and dear by
physicists, such as reproducibility of experiments.

In summary, we will discuss three broad categories of world views
that physicists may hold, which we label fundamentalist,
mystic, and secular. They differ in their view of whether a physics
framework will ultimately give access to all of reality.  The three
answers are, respectively: yes, the current framework; yes, a future
framework; no, no framework present or future.
In what follows, the fundamentalist view is
advocated by Max Tegmark, the secular view by Mark Alford, and the
mystic view by Piet Hut.

A useful starting point for the debate is the Matter-Mind-Math 
triangle (Fig.~\ref{fig:mmm}), put forward by Penrose \cite{Penrose,Penrose2}.

This figure encapsulates the idea that
matter somehow embodies mathematics, the mind arises from
matter, and mathematics is a product of the mind.  If such a circular
production seems unsatisfactory, we can ask which
of these three arrows should be removed.  
In Section~\ref{sec:weak} we see that the fundamentalist, secular and 
mystic viewpoints each find different parts of the diagram to be 
the weakest, and we set forth their arguments and counter-arguments.
This sets the stage for  Section~\ref{sec:beyond}, where each viewpoint
gives its version of how the diagram should be.
Their strengths and weaknesses are debated vigorously
in Sections \ref{sec:Mysticdisc}-\ref{sec:Fundamentalistdisc}.
We close with the three different visions
of the future in Section~\ref{sec:out} and a brief 
conclusion in Section~\ref{sec:concl}.

\section{Three critiques of the Penrose diagram}
\label{sec:weak}

\subsection{The Fundamentalist Critique}

\begin{figure}
\begin{center}
\includegraphics[width=3in]{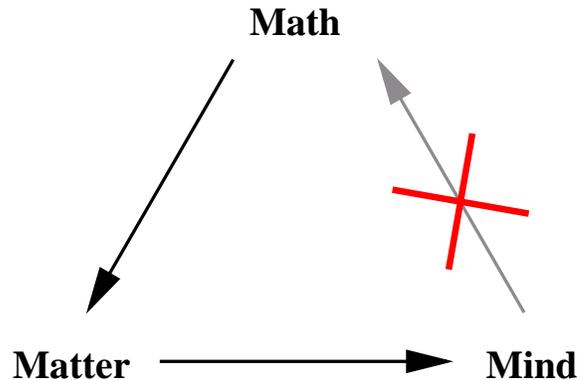}
\end{center}
\caption{Fundamentalist critique of the Penrose diagram}
\label{fig:nonFundamentalist}
\end{figure}

I am a mathematical fundamentalist: I single
out math as the underlying
structure of the universe, and disagree strongly with the
symmetry between math, mind and matter
that is expressed in the Penrose diagram. I have no problem with
the reduction of the world around us, including our minds,
to mathematical laws of physics --- rather, I find it elegant and beautiful.
I am therefore
happy with the Math$\to$Matter and Matter$\to$Mind links, but
object to the Mind$\to$Math link.

\subsubsection{Against Mind $\to$ Math Link}
\label{sec:FundamentalistPlatonic}

I adopt the formalist definition of mathematics: 
it is the study of formal systems.
Although this pursuit itself is of course secondary to the human mind, 
I believe that the mathematical structures that this process uncovers
are ``out there'', completely independently of the discoverer.
Consequently, math is not a product of the human mind, and there
should be no Mind$\to$Math link.

Math is also unique in its ability to
circumvent the classic problem of infinite regress, where every
explanation of a statement in human language must be in the form 
of another unexplained statement. The trick is the emergent concepts 
idea of section 2.E in \cite{Tegmark}.
Although whims of human fashion influence the choice of which
particular formal systems we explore at any one time, and which aspects
thereof, we are continually increasing the amount of charted
territory. 
The street map of mathematical structures is ``out there'', 
and any intelligent entity who begins to study any corner of 
it will inevitably discover at least the main plazas and connecting 
boulevards, even if many charming back alleys and sprawling
suburbs are missed due to cultural prejudice.
The key is that the explorer needs no a priori explanation of what
concepts like integers, vectors or groups mean, since she herself will introduce notation
for them and create her own interpretation of them.
Mathematics is thus the same whether it is discovered by
us, by computers or by extraterrestrials. 

\subsubsection{Defense of Math $\to$ Matter Link}
\label{sec:MathMatterDefense}

Although I respect the Secular ``shut-up-and-calculate'' attitude
(Section \ref{sec:secularcritique})
I feel that the evidence favors the Math$\to$Matter Link.
Physical theories are considered successful if they make
predictions that are subsequently verified.
The view that the physical world is intrinsically mathematical 
has scored many successes of exactly this type, which in my opinion increase
its credibility. 
The idea that the universe is in some sense
mathematical goes back to the 
Pythagoreans, and appears again
in Galileo's statement that the Universe is a grand book
written in the language of mathematics, and in Wigner's discussion of the
``unreasonable effectiveness of mathematics in the natural sciences''
\cite{Wigner}.
After Galileo promulgated the idea,
additional mathematical regularities beyond his wildest dreams were
uncovered, ranging from the motions of planets to the properties of atoms.
After Wigner had written his famous essay, 
the standard model of particle physics 
revealed new ``unreasonable'' mathematical 
order in the microcosm of elementary particles,
and my guess is that history will repeat itself again and again.
I know of no other compelling explanation for this trend
than that the physical world really is completely mathematical,
isomorphic to some mathematical structure.

Let me briefly elaborate on what I mean by this hypothesis that mathematical and physical existence 
are equivalent. It can be viewed as a form of radical Platonism,
asserting that the mathematical structures in Plato's realm of ideas or 
Rucker's ``mindscape'' \cite{Rucker}
exist in a physical sense. It
is akin to what John Barrow refers to as
``$\pi$ in the sky'' \cite{BarrowPi}, what Robert Nozick called the
principle of fecundity \cite{Nozick} and what David Lewis called modal realism \cite{Lewis}.
However, it is crucial to distinguish between two ways of viewing a physical theory: 
from the outside view of a physicist studying its mathematical equations, like a bird
surveying a landscape from high above it, and from the inside view of an observer living in the world
described by the equations, like a frog living in the landscape surveyed by the bird.  
From the bird perspective, the physical world is 
a mathematical structure, an abstract, immutable entity existing outside of space and
time. If history were a movie, the structure would correspond not to a single frame of
it but to the entire videotape. 

Consider, for example, a world made up of pointlike
particles moving around in three-dimensional space. In four-dimensional spacetime -- the
bird perspective -- these particle trajectories resemble a tangle of spaghetti. If the
frog sees a particle moving with constant velocity, the bird sees a straight strand of
uncooked spaghetti. If the frog sees a pair of orbiting particles, the bird sees two
spaghetti strands intertwined like a double helix. To the frog, the world is described
by Newton's laws of motion and gravitation. To the bird, it is described by the geometry
of the pasta -- a mathematical structure. The frog itself is merely a thick bundle of
pasta, whose highly complex intertwining corresponds to a cluster of particles that
store and process information. Our universe is of course far more complicated than this example,
since we do not yet know to what mathematical structure it corresponds. 
Part of the challenge here is that
reality can appear dramatically different in frog and bird perspectives, and the phenomenology linking them 
can be more difficult than finding the correct mathematical structure itself.
It took the genius of Einstein to realize that frogs living in Minkowski space would perceive 
time to slow down at high speeds, and that of Everett to realize that 
a single deterministically evolving quantum wavefunction in Hilbert space contains within it a 
vast number of frog perspectives where certain events appear to occur randomly.

\subsubsection{Defense of Matter $\to$ Mind Link}

I believe that consciousness is the way information feels when being processed.
Since matter can be arranged to process information in numerous ways
of vastly varying complexity, this implies a rich variety of levels and types of 
consciousness.
The particular type of consciousness that we subjectively know is then
a phenomenon that arises in certain highly complex physical systems that 
input, process, store and output information.
This implies that there is nothing wrong with the Matter $\to$ Mind link.

This hypothesis has clearly not been proven. However, this can
hardly be held against it, since it can strictly speaking 
never be proven: I cannot even prove to my colleagues that I personally 
am self-aware---they simply have to take my word for it. Moreover, 
the spectacular successes of computers and neural networks over 
the past decades have arguably made the hypothesis appear less 
implausible than before.

Although believing in this hypothesis may make some people feel less happy
about what they are, I have so far seen no hard scientific evidence 
against it. Rather, many objections seem to be based
on a combination of human vanity and wishful thinking.
Humanity has already had its collective ego deflated so many times
(by Copernicus, Darwin, Freud, infinite Universe cosmology, 
Deep Blue, etc.) that yet another demotion would not bother me at all.
I am what I am and will continue to enjoy feeling the way 
I subjectively feel regardless what the underlying explanation
turns out to be.

\subsection{The Secular Critique}
\label{sec:secularcritique}

\begin{figure}
\begin{center}
\includegraphics[width=3in]{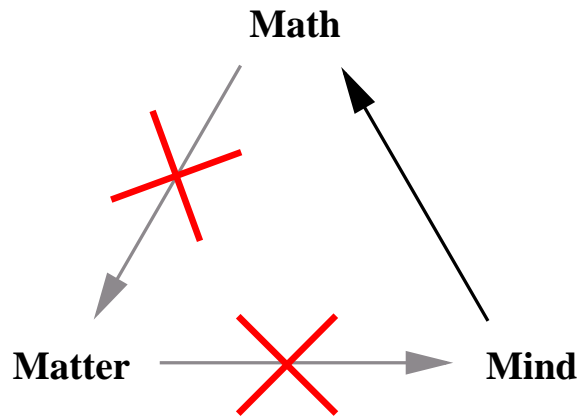}
\end{center}
\caption{Secular critique of the Penrose diagram}
\label{fig:nonSecular}
\end{figure}

I am a secular scientist. I enjoy
practicing science, and believe that it is of great practical 
importance and intellectual value. Everyone should take science
seriously, while remembering that
it is a human creation, not an all-embracing metaphysics.
I am therefore happy with the Mind$\to$Math link.
I appreciate the scientific interest and usefulness of
physiological explanations for human behavior, but
I am unenthusiastic about the Matter$\to$Mind link as a metaphysical
claim. I disagree strongly with the Math$\to$Matter link.

\subsubsection{Against Math $\to$ Matter Link}

All physical scientists are impressed by the ``unreasonable effectiveness
of mathematics in the natural sciences'' \cite{Wigner}.
That effectiveness is indeed spectacular. But
the Math$\to$Matter linkage in the Penrose diagram tries to go further,
inferring from the effectiveness of math that it is 
itself the ultimate substance of the universe.
Wigner, however,
avoiding metaphysical speculation, simply concluded that
we should be grateful for the effectiveness of mathematical methods,
and get on with the business of exploiting them. I agree with him.

I will give my detailed criticism of the Math$\to$Matter link
in section \ref{sec:sec_vs_fund}, where we discuss the
Fundamentalist's version of the MMM diagram, in which this link
plays the foundational role. For now I just want to point out that
although our current theories of physics and our current way
of doing mathematics may seem inevitable, we should be wary of
assigning them any kind of fundamental status. History shows
that both physics and math have changed markedly over time.

The Fundamentalist suggests that math is
a particularly secure foundation because
it can resist the infinite regress of explanation
via ``emergent concepts''.
But an infinite regress of explanations only arises when one is confronted
by the ultimate skeptic, who demands proof of everything.
Such a person will not believe the ``emergent concepts'' claim either,
since it cannot be proved.
For any normal person, explanations will be accepted at some point,
the infinite regress will never arise, and math will not
be needed as a foundation for knowledge.

\subsubsection{Against Matter $\to$ Mind link}
\label{sec:antimaterial}

Like the Mystic, I am unenthusiastic about
the Matter$\to$Mind link, if it is taken as a form of reductionism
that states that the
mental phenomena are ``really just material processes'', as if the
material were somehow more fundamental than the mental. Of course
it is conceivable that everything could be reduced to physics,
but it is also conceivable that it could not. Our scientific theories
are not definitive accounts of nature's secrets, so there is no point
in worrying about what it would be like if they were.
And what happens as our theories of the material world become
better and better? Is there a gradual reduction of the mental
to the physical? Not at all.
We can find out as much as we like about
the brain, but this does not provide a new and better foundation
for our concepts of mental things \cite{Rudd}.
For example, science may
find brain processes that correlate with mystic experiences, 
or genes that determine shyness.
Such progress is extremely interesting, but it happens not
to tell us anything about the {\em meaning}\/ of
``the mystical''  or ``shyness''. 

Of course, scientific progress does influence our beliefs and attitudes.
Rather than saying that our friend is possessed by demons
or went crazy we may now
talk about whether he is suffering from paranoid schizophrenia or
manic depression. These are not just synonyms for ``crazy'', they
represent a richer picture of the person involved, a more specific set
of assumptions about his probable behavior, even a different view of
his curability and humanity.  Our mental concepts have evolved, but
they have not been reduced to physical ones. Even if, as the
Mystic fears (section~\ref{sec:mystic_critique}), we started
talking in materialistic terms---``My brain processes are causing me to
leave the room''---would that mean we were now right where we had been
wrong before?

I therefore find myself sharing
the Fundamentalist's professed indifference to scientific discoveries
about how the brain works.  Would people
feel diminished if it became possible to build Turing-test-passing
machines? I don't think so.  The machines we
currently make are stupid. If we made dazzlingly intelligent, kind and
beautiful machines we would see nothing wrong in being like them.
We would probably claim they were patterned after our image.

\subsubsection{Defense of Mind $\to$ Math Link}
\label{sec:QP}

The Fundamentalist (Section~\ref{sec:FundamentalistPlatonic}) and the Mystic
(Section~\ref{sec:MysticTranscendence}) both express the view that
Math cannot be dependent on the human mind because mathematical
correctness has a solid ``out-there'' feeling about it. 
I do not find this a compelling argument. 
Many conceptual systems with which we are deeply familiar
give us that feeling of inevitable rightness. For example, the
grammar of ones native language has exactly this character: one
``stumbles over'' the incorrectness of ungrammatical sentences.
Even morality has a solid core of indisputably right and wrong
types of actions to which we have an immediate and visceral response, 
though we normally do not discuss these because we take them
for granted and spend our time arguing over the more ambiguous cases.

The denial of the Mind $\to$ Math link corresponds to the
Mathematical Platonist doctrine that mathematical objects have
an independent existence of their own. It is interesting in the
context of the MMM diagram to note that the main argument for
Mathematical Platonism is the ``Quine-Putnam indispensability argument''
\cite{Quine,Putnam}
which exploits the Math-Matter connection. Quine and Putnam 
point out that mathematics is an essential component of science, as are
other exotic entities such as electric fields and protons.
So if we grant that electric fields and protons exist independently
of us, then we should give the same status to the mathematical
objects (real numbers, Hilbert spaces, etc) that we use in science. 
Many philosophical objections have been raised against the
Quine-Putnam argument \cite{StanfordMathPhil}.
As a physicist I find it unconvincing because 
there is a wide range of entities that are posited in physical
theories, from fairly concrete ones like electric fields
to more abstract ones like an object's wave function,
or its center of gravity, that we don't (all) think of as 
``existing'' or ``being real'' in the
normal sense of those words. So just because something plays
a role in our theorizing doesn't mean it is a real object.
And when a physicist argues that something is real
he or she invokes a specific instance where the object
is present:
``of course electric fields are real: you can tell that
there's an electric field around a proton because it pushes
other positive charges away.'' One couldn't speak of
Hilbert space or the complex plane as giving away its
presence in a similar way.
This is related to the fact that what physicists do when they
check whether their theories give a good description of the real
world is very different from what mathematicians do when they
check whether their results are true. In fact, by scientific
standards (and that is what Quine and Putnam
are ultimately appealing to) mathematics really
doesn't have a ``check against nature'' step in its
procedures at all. Checking whether a theorem is true is 
not like an experiment, it is more
like checking whether a sentence is grammatical.

In his critique of the Mind$\to$Math link, the Fundamentalist 
asserts that any
intelligent entity will agree with us about math. For example, aliens
from another planet might have incomprehensible ideas about many
things, but their math would map on to (or extend) ours.
This is a rather vague statement
because it would be up to us to decide which parts of the alien {\em \oe{}uvre}
we call ``math''. It could just reduce to the empty statement that
if we pick out the parts that look like our math, then they will
map onto our math. The Fundamentalist takes it as an article of faith
that a nice mapping, covering a reasonable amount of
material, would necessarily be found, but 
this need not happen. We might be faced with ``inscrutable aliens'',
in whose behavior we cannot cleanly identify anything
that corresponds to ``doing math''.
For example, their successful feats of engineering might not be based on
a coherent mathematical theory of physics but  on
what looks to us like a set of rules of thumb.
Another possibility is ``pragmatic aliens'' \cite{Barrow},
who study mathematical questions,
but with a different approach from our own. 
Like us, they have proven Fermat's last theorem:
they simply computed millions of examples, and declared
it true. Their mathematicians have developed such numerical methods to
a high art, and have numerical ``proofs'' of many theorems that we had not even
conjectured. They seem unimpressed by the rigor of our 
analytic proofs, arguing that
that there is little fundamental
difference between the small chance that their numerical
searches have missed errors in their theorems, and the small chance
that we have failed to notice errors in our analytic proofs.

Of course, even if we find we can agree on
math with all the aliens that we have met, it still wouldn't cast doubt
on the Mind$\to$Math link. It would just mean that in certain respects
our minds are alike. One could speculate on evolutionary reasons
for that, or simply accept it as a fact. And of course, the next
batch of aliens might turn out to be different.

\subsection{The Mystic Critique}
\label{sec:mystic_critique}

\begin{figure}
\begin{center}
\includegraphics[width=3in]{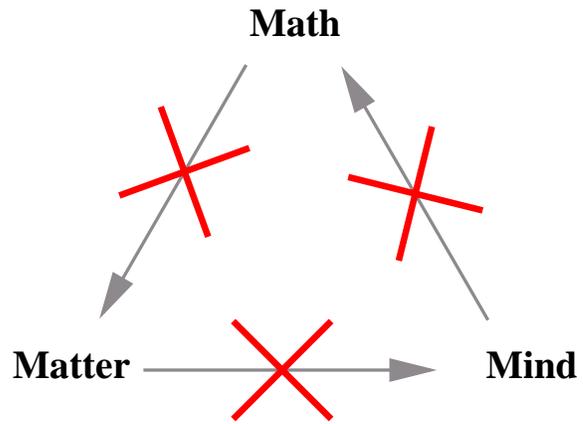}
\end{center}
\caption{Mystic critique of the Penrose diagram}
\label{fig:nonPiet}
\end{figure}

I take a mystic view of science. Unlike the Secularist, I expect that
science will ultimately give us profound insights into the real nature
of the world.  But unlike the Fundamentalist, I believe these will not
emerge in any straightforward way from science as it is currently
constituted.  Rather, I expect science to metamorphose into something
so different that it is literally inconceivable for us.  So in that
sense I agree with the Secularist that physics will probably see
upheavals even (far) more fundamental than the discovery of quantum
mechanics.  And I agree with the Fundamentalist that Science will
ultimately come arbitrarily close to a full understanding of reality.

The reason I like the word `mystic' is that the future science I
envision will be so different from current science, and the role of
elements such as math and experiments will be so different from what
they are now, that we have not the foggiest idea of what these will
look like.  The structures of a future science will remain a mystery,
and the only thing we can be pretty sure of is that our current lines
of reason will be seen to be naive and superficial, compared with the
newer and deeper insights.

So let me be clear: the word `mystic' for me points to a form of
probing into mysteries, as it was meant in Medieval times.  Note
that mystics were very keen to try to show structure and to enumerate
parts of that structure -- the term `mystic' just happened to get a
bad rap later on, and is now unjustifiably associated with attempts
to confuse and muddle a situation.

As for the Penrose diagram, I have deep doubts about all the links.
Making these links now, before a future unification, seems premature.
I strongly believe that the process of unification, which has
successfully uncovered intrinsic links between, e.g., electricity and
magnetism, space and time, matter and energy, will
continue.  What can be more different than matter and energy?  Their
unification was totally unexpected.  If history is any guide, future
unifications will occur that are currently equally unexpected.  And
one example may well involve our three M aspects, matter, mind, and
math.  These three can then no longer can be treated as independent
notions that have the power to point to each other.

Drawing arrows, in my view, is simply a precursor to the program of
unification, in which nature is discovered to be already unified more
than we had thought.  It was through tracing the arrows between
electricity and magnetism -- how exactly can an electric charge
generate a magnetic field, and a magnet generate an electric field --
that electromagnetism was discovered.

Science, like any human activity, is ultimately given in experience,
and understood through the lens of conscious experience.  Within
experience, we can discern subject and object poles.  The trend of
science, so far, has been to explain/reduce all phenomena to processes
that are described purely in terms of objects.  The rise of the
subject is seen as somehow being a byproduct of sufficiently complex
phenomena, taking place in brains, material systems that can be fully
described objectively.  While not denying the correlations between
subjective experience and objective processes in our nervous system, I
do not want to buy into an unquestioned prior status of the object
pole over the subject pole of experience \cite{Hut0}.

The Fundamentalist defends the Matter$\to$Mind link on the basis
of advances in neurophysiology.
I do not deny that a
deep understanding of the material structure of the human brain will
shed a lot of light on the way {\it in which} we experience; but the
very fact {\it that} we experience may completely fall outside such
an explanatory framework.  My guess would be that such a question
requires a shift to a wider horizon of knowledge/meaning/explanation 
\cite{Hut1,Hut2}.

Let me give an analogy.  
Imagine a world in which there are no
periodic phenomena, and hence no clocks of any type in nature.
Someone living in such a world may not easily discover the concept of
time, and certainly not a concept of a linearly progressing time, that
can be mapped onto a one-dimensional line of real numbers.  But it would
be wrong to draw the conclusion: no clocks no time, hence clocks
produce time.  Similarly, it is too simple to say: no brains no
experience, hence brains produce experience.  Reality may be a lot
more subtle.  Brains may tune into an aspect of reality that is
explicitly filtered out when setting up laboratory experiments and
when formulating mathematical regularities summarizing the lab
experiments.  This is just one analogy, and merely meant to illustrate
the fact that there is a lot more room within the ``horizon of knowing,''
than we normally consider.

To try to explain our experiences as somehow arising out of matter is
a tempting project. 
And of course there is a very tight correlation
between the thoughts we think and the precise electromagnetic and
chemical processes in and between our brain cells.  
However, the problem with identifying brain states with experience
is that we are short-changing ourselves.  
When people talk 
about their experience by invoking
scientific images, such as ``my hormones drive me to do such and such''
or ``my taste buds enjoyed so and so'',
they are
using technological thinking as a mode of alienation.

When the Fundamentalist says ``I am what I am'',
it strikes me as hopelessly naive. While a sudden new discovery will not
instantaneously change the way we experience the world, by
the time the new knowledge seeps into the way we view the
world, it definitely colors the way the world presents itself
to us. The existence of placebos is just one striking example
of this phenomenon.

\subsubsection{Transcendence of all three links}
\label{sec:MysticTranscendence}

I neither want to attack nor defend the three links in the Penrose diagram,
in any absolute way.  Rather I want to transcend them, after first defending
them in a relative way.

The proper defense of all three links is not by ascribing to them
any power in terms of causality, but by pointing to meaningful
correlations that exist between the three M-elements.  To take the
example of a movie: within the movie, all kind of phenomena seem to
`cause' other phenomena, in rather precise ways, but we know that the
real cause lies in the projector, and in the process of shooting the
film in the first place.  The correlations are still important, since
without them the movie would be just a heap of flickering lights
without meaning.  But the importance of the correlations is strictly
limited to the framework of the movie, and has no fundamental meaning
whatsoever.

So perhaps this is what makes me a mystic, in that I am uttering
seemingly contradictory statements.  Among the three of us, I am the
only one who can accept all three links, without being bothered by the
circular nature of the `vicious triangle' links.  Since I only ascribe
relative meaning to them, I have no need for any foundation.  None of
the elements within the story of reality is absolute or basic; all
elements emerge simultaneously from a deeper unification, speaking in
physics lingo.  In Buddhist lingo, for example, this could be called
co-dependent arising \cite{Napper}.

The Platonic position of the Fundamentalist, that mathematical truth
exists all by itself, is extremely appealing because it seems to
correspond closely to our experience.  When we struggle with a
mathematical problem, and finally find a solution, we sometimes
``stumble'' upon it in a way that is rather similar to the way we
stumble over a chair.  The resistance that mathematical objects show
to our attempts to prove what later turns out to be false is akin to
the resistance that physical objects show when we try to wish them
away -- both appear to have an existence independent of the presence
or absence of individual humans.  In this sense, I am sympathetic to
Alain Connes' notion of archaic mathematical reality as being as real
as physical reality \cite{CC}.

The Secular view that matter and mind have their own meaningful
existence I also find extremely appealing, within our every-day
explanatory framework.  In practical terms, it makes sense to
deal with the world around us in terms of material objects and
energetic processes, and it also makes sense to treat our experience
as something that has equal pride of place.

The problem arises when we try to isolate elements from this
story, and point to some of them as truly fundamental.  In my view,
a future physics will transcend any story we have woven so far.

\section{Beyond the Triangle: Three Views}
\label{sec:beyond}

\subsection{The Mystic View: the Other-Source Diagram}
\label{sec:Mystic}

\begin{figure}
\begin{center}
\includegraphics[width=3in]{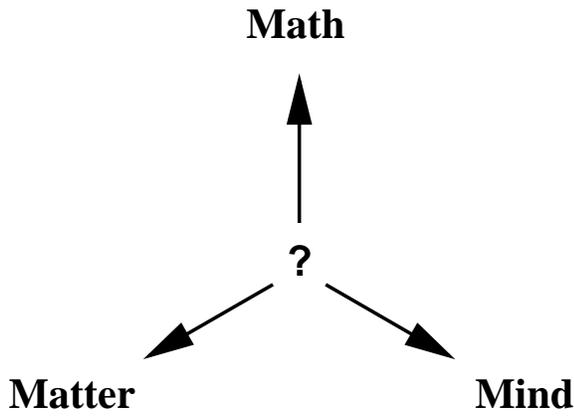}
\end{center}
\caption{Mystic's view: Other-source Diagram}
\label{fig:Mystic}
\end{figure}

\noindent {\bf Mystic:}\\
I'm against all three arrows in the original picture.  In my alternative
picture, ``?'' stands for an origin that cannot be easily described, the way
each of the other three can.  Our three M's are more like the shadows on
the wall of Plato's cave; or in another metaphor, they are the fish
that can be dragged up with the nets of discursive/conceptual thinking.
The Source or Origin lies beyond that, and is more real than 
any particular element of what we conventionally take reality to be.

Note here that in fact, upon finer scrutiny, the separation between ``?''
and the three M's is only illusory.  The real mind cannot be captured
in a description, nor can the real matter.  I'd say that even the real
math cannot, if you include the living intuitive process of discovery.

\subsection{The Secular View: the No-Source Diagram}
\label{sec:Secular}

\begin{figure}
\begin{center}
\includegraphics[width=3in]{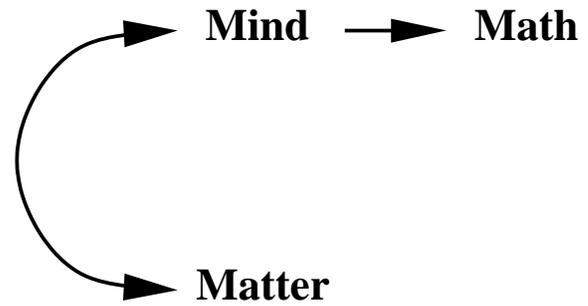}
\end{center}
\caption{Secular view: No-source Diagram}
\label{fig:Secular}
\end{figure}

\noindent {\bf Secularist:} \\
In so far as the relationship of matter, mind, and math can be
expressed in the form of Penrose-type diagrams, I prefer figure
\ref{fig:Secular}. It presents mind and matter as two
different concepts, but ones that are used together in a single
picture of the world.
It places math at a lower level than matter
and mind: math is a particular mental activity, among many others.
There is no reason to give it
a privileged position above, say, morality or language use.


Although mind and matter are separate in this diagram, I do not want to
imply that they are disjoint (dualism). I see them as two aspects
of the world that we happen to distinguish quite sharply.

\subsection{The Fundamentalist View: the Radical Platonist Diagram}
\label{sec:Fundamentalist}

\begin{figure}[t]
\begin{center}
\includegraphics[width=3in]{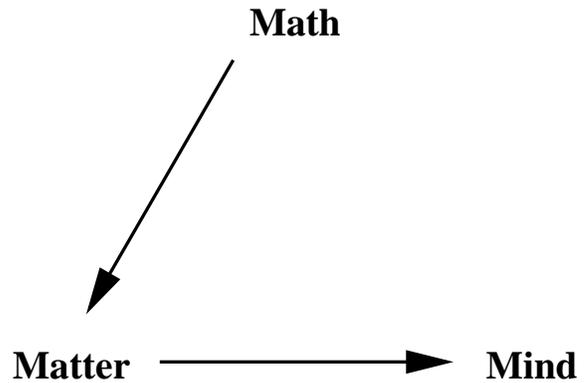}
\end{center}
\caption{Fundamentalist view: Radical Platonist Diagram}
\label{fig:Fundamentalist}
\end{figure}

\noindent {\bf Fundamentalist:}\\
My view is that mathematical structures 
(the cube, manifolds, operator algebras, etc.)
exist quite independently of us humans, so
math must be promoted to the fundamental vertex, as in
figure~\ref{fig:Fundamentalist}.
The human mind then emerges from math, as a self-aware substructure
of an extremely complicated mathematical 
structure \cite{Barrow,Tipler,Schmidthuber,Tegmark,TegmarkMultiverse}.
Each such substructure 
subjectively perceives itself as existing in a physically 
real sense. 
Given the mathematical equations that describe our Universe,
an infinitely intelligent mathematician could in principle 
deduce the properties of both its material content and the
minds of its inhabitants.

\section{Debating the Mystic's Other-Source diagram}
\label{sec:Mysticdisc}

\subsection{Secular Critique of Mystic's Other-Source Diagram}
Firstly, I would say that the main difference between the 
Mystic and the Fundamentalist
is in the question of timing. The Fundamentalist thinks we already
know the ultimate constituents of the world; the Mystic thinks we will
come to achieve such enlightenment in the future. This makes it harder
to criticize the Mystic's approach, because it does not yet have
any specific form, but the basic objection is the same: why will we
be entitled to declare the final victory at some particular moment?
When is the right time to promote whatever picture we have found
workable to metaphysical status?

Secondly, I do not think it is meaningful to invoke a hidden or
ineffable source.  If it cannot even be described, what role
can it play in our quest for knowledge? Playing with the
definition of the word ``real'' seems to me to be a similarly
arbitrary exercise.

Thirdly, I do not think that math deserves to be put on the
same footing as mind and matter. There are non-mathematicians, but
there is nobody who is non-mental or non-material!
What determines which human activities we include as separate poles
connected directly to the ineffable origin? Why not include
language, music, art, and so on?

\subsection{Mystic's Response} 

I see science as converging further and further toward what is true
about reality, but I don't expect there to be a particular point at
which we can declare victory.  The road toward deeper insight may
be unending, simply because the degrees of what can be called insight
may be inexhaustible.  Classical mechanics seemed to give complete
insight, in principle, but quantum mechanics offered a type of insight
that was qualitatively deeper and other.  I have no reason to expect
this process of discovery to stop.  My main point is not that we will
reach a final truth, but rather that we will never reach an absolute
boundary beyond which we cannot progress in getting closer to the truth.

As for the ineffability of a source, that is just a relative notion.
When all you have are concepts such as electricity and magnetism, then
electromagnetism is ineffable, since it doesn't fit into your framework.
It has to be invoked as a new type of source that can be projected in
your existing framework, very much like the shadows in Plato's cave.
But then you enlarge your framework in order to include the new source.

By the way, the requirement that any source should be describable is
not something I want to buy into.  Just as quantum mechanics has
tossed out some of the deepest held convictions of classical mechanics,
similarly I expect a future science to throw out some of what we now
still see as absolutely essential for the scientific method.  Our current
notion of a description may be one of them.  We already see a harbinger
of this shift in quantum mechanics, where the perfectly effable classical
idea of an exhaustive description of all degrees of freedom of a particle
has turned out to be impossible.

As for the Secularist's existence proof for non-mathematicians,
I agree.  However, I had in mind something wider than technical
mathematics.  If we want to classify what we find around and in us under
the three headers of matter, mind, and math, we'd better take each of
these terms rather generally.  Under matter we take anything physical,
such forms of pure energy like radiation, or even the vacuum
nowadays.  Similarly, under math we should take any conceptual
structure that we use in our mind in order to make sense of both mind
and (material) world.  Language, logic of all sorts (including common
sense reasoning), any distinctions that are found to be useful I would
group here under ``math'' -- take counting, for example, as a mapping
operation to divide things in the world into distinct categories.
In this way, I think you will agree that there are no non-conceptual
humans, as long as they function normally.

In other words, I don't think we are dealing with mathematics here as
just one of out many academic subjects, such as physics or biology or
art or literature.  The first two subjects 
apply more to the ``matter'' corner of
our triangle, the latter two more to the ``mind'' corner of our triangle,
but ``math'' stands out as the surprising fact of the regularity of the
world, or if you like, the surprising fact that we can view the world
in such unexpectedly regular ways and get away with it, without
contradictions, and with the ability to make accurate predictions.

\subsection{Fundamentalist Critique of Mystic's Other-Source Diagram}

My main objection to both the other source view and the no-source view
is an admittedly emotional one: I perceive them as defeatist.
Figures~\ref{fig:Mystic} and \ref{fig:Secular}
both indicate that the quest for a fundamental theory of everything is 
ultimately doomed, and that the best we can do is go
off and meditate or optimize a ball bearing.  
Although the Mystic hopes for great conceptual breakthroughs ahead,
I am not persuaded by his hypothesis that we will come arbitrary close to a complete understanding while 
remaining precluded from ever attaining it.
I also find the Mystic battle plan disturbingly vague, failing to understand what specific steps we should take 
to explore the object pole.
In contrast,                                                                                                                                                                                                                                                                                                                                                                                                                                                                                                                                                                                       
Figure~\ref{fig:Fundamentalist}
can be taken as a specific battle cry to forge ahead in search of the equations
of a truly fundamental theory. This theory may turn out to be too
complex to be comprehensible to us humans, but at least there's hope.
We shouldn't give up without trying!

\subsection{Mystic's Response:} 
I applaud the
enthusiasm of the Fundamentalist, and I share it.  I only differ in the
direction in which I think we are likely to find the answers. 
The Fundamentalist is in effect looking for the lost key under
the lamp post.  He takes the objectivistic program, in which
everything is explained in third-person terms as interacting objects,
and hopes that that program will carry the day.  But I see that as a
limiting case of a wider program, in which first-person experience
and a study of the subject {\it qua} subject will augment our already
very detailed studies of objects.
Within that wider program, I think we have much more of a chance to
find a horizon in which we can see how all of our experiences hang
together in a meaningful way, without artificially reducing everything
to a particular subset of privileged phenomena.

What we really need to do is to go back to the state of science as it
was in the days of Galileo.  He focused on some very simple experiments,
in which he rolled balls from inclined planes and the like.  I think
we should similarly focus on very simple experiments in which we take
the subject equally seriously as the object, and try to see where that
leads us.  Such an approach may well point the way to deeper forms of
unification, including the unification of subject and object \cite{Hut0}.

\section{Debating the Secularist's No-Source diagram}
\label{sec:Seculardisc}

\subsection{Fundamentalist Critique of the Secularist's No-Source Diagram}
The Secular view
seems even more defeatist to me than the Mystic one.
In the Mystic's view, there is at least an underlying unity that we may
one day be able to understand better. The Secularist
doesn't even want to bother searching for an underlying reality,
since it simply doesn't exist. We are admonished to stop seeking a 
deeper understanding since there is nothing deeper to understand. 
I am convinced that there is a deep explanation for it all,
the only question being whether we can understand it 
from the frog perspective of our limited human minds.

\subsection{Secular Response} 
The Fundamentalist needs to open his eyes to the world that
lies outside theoretical physics. Of course there is a thrill in the
thought that your research is opening up the blueprints of
the universe, but it is the guilty pleasure of surrendering to
vanity and parochialism. To submit to it, you
must convince yourself that we are living at a special time,
when the basic structure of math has been determined (``formal
systems''), and the final theory of physics is at hand. These
metaphysical yearnings are easy to empathize with, but
the whole approach is shamelessly ahistorical. 
It has beneficial side-effects (the ability to raise public
enthusiasm, for example) but also leads to a devaluation of the
normal process of day-to-day science.

The Fundamentalist says we should be searching for an ``underlying
reality''. I agree that the laws of nature are not immediately
obvious, but reality itself is not lying under anything.
It is available for us to explore by scientific or other
means.  Scientific methods have been highly successful in
giving us a coherent picture of and control over the world we live in.
In any
reasonable sense of the word ``understand'', I believe there are many
aspects of the world that we either already understand or can hope to
understand in the future.  However, the
Mystic and Fundamentalist both want to go
further, to an {\em unimprovable} understanding.  They have some concept of a
hidden universal truth. But this can only be defined as that
which is yielded by the method
they suggest for finding it, and there is no reason to think
that this method will converge to a unique result.

\subsection{Mystic's Critique of the Secularist's No-Source Diagram}
Even though the Secularist claims not to be a dualist, 
and affirms unity between
the aspects of mind and matter, his description of mind and matter as
two different ``aspects'' makes the two so separate as to make it very
hard for me to see where the unity lies.  At least a material
reductionist has a clear form of unity to show for, in that everything
is considered to be reducible to (properties of) matter.  In the
Secular view, the two ``aspects'' that we 
``happen to distinguish sharply'' form
an effectively dualistic picture of the world.

The whole notion of dualism as having any ultimate standing, 
explicit or implied, goes against
all that we have learned in the history of physics.  Time and again, 
we started studying two seemingly separate topics, like space and time,
or electricity and magnetism, or matter and energy, only to find out that
they were merely poles or aspects of a single
more fundamental entity, such as
spacetime, electromagnetism, or matter-energy. In other words, physics
teaches us that dualistic descriptions are nothing but harbingers of a
deeper level of non-dual understanding.

Even apart from my intuition as a physicist, all I ever see in 
dualistic thinking is some form of either
laziness or fear.  Yes, it is much simpler to keep the world neatly
partitioned, and then to study either side of the fence at leisure.
Crossing the fence requires more work and is more risky, in that it
may well undermine, at either side of the fence, many prejudices held
near and dear for a long time.

Specifically, the strongest argument against dualism is: if mind and
matter really have nothing to do with each other, how come they show
up in the same world, and how come they do {\it seem} to interact so
tightly?


\subsection{Secular Response} 
Perhaps the Mystic and Fundamentalist are reading too much into my
diagram: it is drawn in a pragmatic rather than a metaphysical
spirit.  By their standards I should refuse to draw any diagram at
all.
I see my diagram as similar to the common-sense one
that divides living beings into plants and animals. These are
obvious categories that reflect important distinctions.
That is all it is trying to say.

\section{Debating the Fundamentalist's Radical Platonic diagram}
\label{sec:Fundamentalistdisc}

\subsection{Secular Critique of the Fundamentalist's Radical Platonic Diagram}
\label{sec:sec_vs_fund}

The Math$\to$Matter link plays a crucial role in the Fundamentalist's
diagram. It corresponds to a very strong set of statements:
\begin{list}{}{
  \setlength{\itemsep}{-0.7\parsep}
  \setlength{\topsep}{-0.7\parskip}
 }
\item[{\bf (F1)}]Physics is converging to some set of ultimate laws,
\item[{\bf (F2)}] Our study of mathematics is converging towards an 
    ``ultimate math''.
\item[{\bf (F3)}]The ultimate laws will be expressed
  in terms of the ultimate math.
  Thus math has a special relationship to the material world.
\item[{\bf (F4)}] It is plausible (or perhaps just aesthetically appealing) 
 that the ultimate math is not just an external reality, but that
 our Universe consists of it.
\end{list}
I find all these claims to be either meaningless or dubious.

\noindent 
\underline{Against (F1)}:\\
It is not clear that physics is converging to a set of ultimate laws.
Physics has changed greatly even in the last 100 years.
Our basic understanding of space and time was revolutionized, and even
the core concept of the determinism of physical law was radically
revised by quantum mechanics, which introduced elements of randomness
that classical physicists would have found unbelievable. There is
no reason to think that physics is now immune to such upheavals, or
that it ever will be.
A related question is: when will we declare 
that the convergence is complete? The cost of testing our
``fundamental'' physical theories
has escalated dramatically over the last
few decades as the relevant energy scale rises. 
It seems likely that as these theories are successively
improved, it will become harder
and harder to do the experiments that would test them. 
The theory that survives
uncontradicted may do so mainly because the difficulty of falsifying it
becomes insuperable. Another way to say this is that the time between
revolutions in physics will become longer,
and at some point one will not be able to tell whether
it is infinite.  When this happens, the human need for novelty
will lead science down other, more rewarding, paths than
the search for a fundamental theory.

\noindent 
\underline{Against (F2)}:\\
Math is certainly progressing, but there
is no reason to think that it is converging to a definite
structure, with fixed questions and established ways to address them.
Consider the history of math. There was mathematics
before Hippasus the Pythagorean showed that $\sqrt{2}$ is irrational;
mathematics before Russell's paradox, when it was thought
that arithmetic could be reduced to set theory; 
the famous battle between intuitionists and formalists,
before G\"odel showed that there were true but undecidable
propositions; the computer proof of
the 4-color theorem by Appel and Haken.
In each case we find that progress did not just fill in some
gaps, but modified the idea of what mathematics is or
should be. Mathematics has taken radical turns in the past
and will do so in the future,
stepping out in new directions that confound attempts to
systematize it once and for all.
I think this makes it impossible to formulate a
coherent concept of the ``ultimate mathematics''.

\noindent 
\underline{Against (F3)}:\\
It is striking that mathematical methods have been outstandingly
successful in physics. But what is the content of the assertion that
any supposed ultimate laws of physics must be ``mathematical''?
As I argued above,
mathematics has evolved over time. It has also been strongly influenced by
physics. Five hundred years from now physicists may be using methods
that we would find hard to recognize, and mathematics may
have grown to include those methods simply because they work for physicists.
Or perhaps the intellectual map will have changed to the point where
we can't cleanly identify a domain that we would call ``physics''.
Perhaps the Fundamentalist would prefer to  claim that
ultimate physical laws will have the character of what we {\it now}
call mathematics. That is simply a statement of faith, and
the Fundamentalist does not offer any reason to believe it.

\noindent 
\underline{Against (F4)}:\\
(a) Plausibility. 
The Radical Platonist position 
is just another metaphysical theory
like solipsism, and even materialism.
In each case one starts with a decent theory (mathematical
descriptions of nature, sense-data as the basis of cognition) and then
raises it to an object of worship, as having finally captured the
essence of reality.  Philosophers from Kant to Wittgenstein have
criticized such thought for taking loose use of language to an extreme
where it fosters intellectual illusions \cite{Hacker}.  In the end the
metaphysics just demands that we use a different language for saying
what we already knew. But what is the difference?
Nothing is gained by reformulating ``it is
dark in here'' as ``I have few visual sensations'' or ``the local
intensity of electromagnetic radiation is low''.  Talk of
conscious beings as ``self-aware substructures'' is a similarly
empty transcription.

%
\noindent (b) Aesthetic appeal.
It is not clear that Math $\rightarrow$ Matter is aesthetically
attractive.  Even if there were some ultimate mathematics, it might be
nothing like what we {\it now} call math.  It could easily seem as
ugly and contrived to us as irrational numbers and Russell's
``repaired'' set theory did to their contemporaries when they were
formulated.

Finally, the Fundamentalist describes himself as a formalist and a
Platonist, but these are contradictory. A Platonist believes
mathematical truths are truths about some world of mathematical
objects, while a formalist believes that math is just the sum of all
strings that you can get by manipulating symbols according to rules,
starting with arbitrary axioms.  As G\"odel's theorem shows, these are
two different things: the methods allowed by formalists cannot prove
all the theorems in a sufficiently powerful system. 
There are systems as powerful as
arithmetic that are consistent and complete, and that therefore {\em
cannot be axiomatized} (Ref.~\cite{Barrow}, p.~126), 
and so are outside the formalist structure.
This spells doom to the Fundamentalist's project.  The idea that math is
``out there'' is incompatible with the idea that it consists of
formal systems.

\subsubsection{Fundamentalist reply}

I would rephrase my assumptions as a Fundamentalist as follows:
\begin{itemize}
\item {\bf Assumption A1:} That the physical world 
(specifically our Level III multiverse \cite{TegmarkMultiverse}) 
is a mathematical structure.
\item {\bf Assumption A2:} Mathematical multiverse:
that all mathematical structures  exist ``out there'' in the same sense.
\end{itemize}

The Secularist (F1) critique of A1 is focused on the question of whether 
our understanding of physics will converging towards fundamental laws. 
I believe that if we fail in this quest, if will be because of the
limitations of our human minds rather than because of the nature of reality: 
I view it as almost tautological that there are some fundamental 
laws that nature obeys:
I assume that there is an external reality that exists independently of us humans,
and the laws of physics are how this reality works.
Denying this external reality would be flirting with solipsism, 
and I view it as human vanity taken to the extreme.

Although the (F2 \& F3) critique above suggests that ``ultimate math'' is vague
and undefined, there is nothing vague about my two assumptions.
The notion of a mathematical structure is rigorously defined in any book
on Model Theory.  The integers are well-defined even though most of
them have never been used in human calculations, and mathematical
structures are likewise well-defined even though most of them have yet
to be explored by mathematicians.  
The Secularist argued in Section~\ref{sec:QP} that
alien mathematics might be unrecognizable to us.
If so, it would only be because we
are uncovering a different part of what is in fact a consistent and
unified picture, so math is converging in this sense.
The reason why this is far from apparent is that
our development of mathematics is in a very early stage, nowhere near
a systematic classification of even the most basic 
structures/formal systems. Our
attention is therefore drawn to interesting features in the
mathematical landscape such as theorem $X$ or formal system $Y$, and
we may not yet see the forest for all the trees.


The secularist criticizes (F4) for loose use of language.
To me, {\it all} use of human language is necessarily loose and hence 
insufficient for describing an external reality existing independent of us.
This is why nature speaks the language of mathematics and 
why I am advocating mathematical language to describe reality.

The objection that (F4) may be aesthetically unappealing 
takes the anthropic principle to a comical extreme, suggesting that the universe
must be devised so as to make us like it (note also that by Assumption 2, we are part of merely one
particular mathematical structure in the vast Level IV multiverse of
all structures, some subjectively more elegant than others).
I view the mystic's objection in Section~\ref{sec:mystic_critique} to technological thinking as a mode of alienation
as either invoking similarly wishful thinking
(``our universe must be devised so as to not make us feel alienated'')
or as encouraging thought control.

Aside from Wigner's above-mentioned ubiquity of mathematics in
physics, a second argument supporting assumption A1 (and F3) is that abstract
mathematics is so general that {\it any} fundamental ``theory of
everything'' that is definable in purely formal terms (independent of
vague human terminology) is also a mathematical structure.  For
instance, a TOE involving a set of different types of entities
(denoted by words, say) and relations between them (denoted by
additional words) is nothing but what mathematicians call a
set-theoretical model, and one can generally find a formal system that
it is a model of.  In other words, if the physical world exists
independently of us humans, it is not obvious that it can avoid being
a mathematical structure.


Given A1, a second argument for assumption A2 (and F4) is that if two entities are
isomorphic, then there is no meaningful sense in which they are not
one and the same \cite{Cohen}.  This implies assumption A2 when the
entities in question are a physical universe and a mathematical
structure describing it, respectively.
To avoid this conclusion that mathematical and physical existence are
equivalent, one would need to argue that our universe is somehow made
of stuff perfectly described by a mathematical structure, but which
also has other properties that are not described by it. 
However, this
violates assumption A1 and implies either that it is isomorphic to a
more complicated mathematical structure or that it is not mathematical
at all.  The latter would be make Karl Popper turn in his grave, since
those additional bells and whistles that make the universe
non-mathematical by definition have no observable effects whatsoever.


Finally, I find the objection involving G\"odel to be very interesting
and subtle.  My hypothesis is that only G\"odel-complete (fully
decidable) mathematical structures have physical existence.  This
drastically shrinks the Level IV multiverse, essentially placing an
upper limit on complexity, and may have the attractive side effect of
explaining the relative simplicity of our universe.  If you define
mathematical structures as (equivalence classes of) models of axiom
systems \cite{Tegmark}, then they are guaranteed to be
G\"odel-complete (consistent).  Please note that although we
conventionally use a G\"odel-undecidable mathematical structure
(including integers with Peano's recursion axiom, etc.) to model the
physical world, it is not at all obvious that the actual mathematical
structure describing our world is a G\"odel-undecidable one ---
lacking a theory of quantum gravity, we have certainly not found it
yet. Even a world corresponding to a G\"odel-complete mathematical structure could 
in principle contain observers capable of thinking about G\"odel-incomplete mathematics,
just as finite-state digital computers can prove certain theorems about 
G\"odel-incomplete formal systems like Peano arithmetic.

The classification project for consistent mathematical structures may
well prove far too difficult for us humans to complete, but even
partial success in this endeavor could be useful, since the tiny
fraction of mathematics uncovered so far has gone such a long way in
understanding the physical world. So let's not give up without trying!


\subsection{Mystic Critique of the Fundamentalist's Radical Platonic Diagram}

On the convergence of math, I remain agnostic.  I agree that, historically,
new developments were completely unpredictable, and unexpected.  And yes,
there is a ``resistance'' to mathematical objects that
makes them in many aspects like physical objects.  As I mentioned
earlier, you can stumble on properties of triangles, like you can
stumble on furniture.  In that sense, there are certainly not a matter
of fashion or convention (although the framework, within which there
are defined, may be; I guess that is Secular's main point).
The Fundamentalist's belief that the current framework is more than
fashion is, I guess, a belief.  I don't see clear arguments in favor
of it, although I must say, intuitively, it is
rather appealing. 

I share the Secularist's doubts about the idea that the world is
``made of math''.  My argument against this is that it seems to me like a
category mistake.  The category of existence of physical objects is
different from the category of existence of mathematical objects.  For
one thing, this particular chair here exists, and is distinct from a
``chair in general'.  Triangles always belong to the latter category;
I've never come across a particular triangle with an individual
existence like that of that of a chair.

My intuition tells me that mathematics can never be exhausted.
Whenever we have tried to formalize a system in the past, we wound up
formalizing some neatly delineated piece of turf, very interesting by
itself, but leaving out unexpected other developments that did not fit
the mould.  Unless the Fundamentalist comes up with a very good
argument for the possibility of finding an overarching ``space of all
models'' together with a clear structure governing that space, I don't
see how his project could ever work.  In mathematics, asking such
questions as ``what is the set of all sets'' has always run into
inconsistencies.  I would expect something similar to happen here.

In summary, I love the Fundamentalist project of
classifying mathematics to such an extent that the parallels between
``what is there'' in math and ``what is there'' in physics become
compelling.  I don't expect this project to be possible, in the end, but I
do expect interesting insight to come out of trying to make it work,
so therefore I'm all for it.

\subsection{Fundamentalist Response}
The resolution to the category issue is that the entire Universe 
is a single mathematical structure, subsuming within it all particular physical objects.


I share the Mystic's concern that the Radical Platonist program may be
extremely difficult for us humans to carry though even if the
underlying assumption is correct. However, it's worth a shot! 
My description of my vision for the future (below)
lists a few areas where it should be possible to make
at least limited progress.

\section{Three visions of the future}
\label{sec:out}

\subsection{The Fundamentalist Vision}
If the radical Platonist view is correct, both arrows in 
Figure~\ref{fig:Fundamentalist}
deserve intense study since there is real hope of understanding
them better. This means continuing with
``business as usual'' in both fundamental physics research
and brain/consciousness research. 
However, it also suggests research in some slightly unconventional 
directions as described below.

If all mathematical structures are equally real, then the one we
inhabit is but one in a vast ensemble, and should be expected to be
the most generic one compatible with our existence.  In the
terminology of \cite{TegmarkMultiverse}, this ensemble or ``Level IV
multiverse'' is vastly more diverse than the Level I multiverse of
spatial horizon volumes, the Level II multiverse of different
post-inflationary domains where the same fundamental laws have
produced different effective laws from its landscape of possibilities,
and the Level III multiverse of unitary quantum mechanics.  To test
this prediction, it is interesting to work out how the Universe would
differ if physical constants or equations were altered, quantifying
the degree to which it appears fine-tuned for life.

The ultimate classification problem in mathematics would
be to classify all formal systems. Very little progress has
been made in this direction because of the great difficulties 
involved, but any further insights about the structure of 
mathematics could shed more light on the nature of physics.

As described in Section~\ref{sec:MathMatterDefense},
a mathematical structure can be viewed in two ways: 
from the bird perspective of a mathematician or from the 
subjective frog perspective of a self-aware substructure in it, 
like us. Since the relation between these two perspectives 
can be extremely subtle, a more systematic study of such phenomenology
issues will be important --- otherwise our candidate mathematical theories will not make 
testable predictions, and we may not even recognize the 
correct equations if we stumble across them.

Finally, since all mathematical structures are equally real in this
view, we need to keep a very open mind as to what we are looking for.
There is no room whatsoever for subjective nostalgic bias
towards structures that resemble cozy classical concepts, or rejection
of theories because they are ``too crazy''.

In conclusion, I feel that the fundamentalist vision is the most specific and falsifiable of
the three, subsuming the Secularist's to-do list and adding to it items that are less vague than the Mystic's.
There are real calculations to be done here, and my opinion is that
controversial philosophical arguments should never be used as excuses not to make calculations.

\subsection{The Secular Vision}

Can a Secularist express a vision for the future? The whole
point of the secular view is to avoid grandiose posturing.  I have
already given my expectations about the future of fundamental physics
(Section \ref{sec:secularcritique}): as it becomes harder
to obtain experimental data against which to test our most
fundamental theories, it will become harder to know whether we should
believe them. One cannot
predict how fundamental physics will then progress.  It might
continually find unexpected directions in which to proceed, such as the
one hoped for by the Mystic. It might change so drastically that 
we would no longer recognize it as physics, or even science.
Finally, it might simply atrophy away.

If I grapple with any deep mysteries, they are not about the meaning of
science, but about the way people see the world. Science is the process of
building a picture of the world that is, where possible,
quantitatively accurate. It strikes some balance between conservatism
(making minimal changes to established pictures) and radicalism
(abandoning pictures that do not work). But many people feel that
science is incomplete without a deeper claim than mere predictive
power.  They want to tilt the balance towards conservatism, treating
the picture, current or future, as metaphysical truth.  The mystic
and fundamentalist thinkers show this tendency, and, in a wider
context, some of the most successful religions
display an extreme form of such conservatism,
requiring adherence to a rigid set of beliefs that vary very
slowly. It is striking, and mysterious to me, that so many people feel
a strong need for such a frozen picture.

What impresses me is the degree to which our current understanding
of the universe would be utterly incomprehensible to previous
generations. It seems very reasonable to expect
that  human ideas will continue to develop
in unpredictable ways, and that the theories of the future will
explore directions that we cannot even imagine today.
Ironically, it may be the fundamentalist and mystically inclined
thinkers who bring this about. I am inclined to
believe that those who think that
their theories uncover the deep structure of the world
are the ones who are driven most strongly to make essential and 
revolutionary contributions
to the progress of theoretical physics.


\subsection{The Mystic Vision}
We have painted ourselves in a corner, scientifically, by describing
the whole world in objective terms, and finding less and less room for
ourselves to stand on.  The challenge we now face is not to reduce
ourselves also to objects, but to explore ways to let science
naturally widen its area of investigation, while staying true to its
methodology of peer review, based on an interplay between theory and
experiment, with experiment having the last word.

For the last four hundred years, natural science has studied the
object pole of experience in ever increasing detail.  While this has
so far been a sensible approach, we are now reaching the limits of a
purely object-oriented treatment.  In various areas of science, from quantum
mechanics to neuroscience and robotics, the subject pole of experience
can no longer be neglected.  Most likely, science will change
qualitatively with this required extension of its methodology.

This will not happen overnight.  I expect this program to be carried
out over a period of time comparable to the time it took to get
the science of the object up and running, perhaps a century, quite
likely a few centuries.  But this process cannot and should not be
hurried by wishful thinking or by external agendas.  Imagine what
would have happened if physicists had listened to William
Blake, two centuries ago, who fulminated against their clockwork
picture of the Universe.  If they had tried to start a new
``poetic physics'' by trying to force Blake's notions into the accepted
framework of physics, they would have been led astray.  Nothing has
come, fortunately, from ``communist biology'' in the days of Lysenko or
the ``Aryan physics'' of the Nazis.

Real progress in physics can only come from within, from a
necessity to introduce wider frameworks of explanation and
interpretation to accommodate experimental facts that cannot be
satisfactorily dealt with in the existing frameworks.  
After Blake's complaints, more than a hundred years passed before
physicists discovered quantum mechanics, which showed that the
material world is indeed really a far cry from a clockwork universe.
The original dogmas of repeatability and strict causality, in the
classical sense in which nothing happens spontaneously, were shown to
be incorrect.  These ideas had been extremely helpful to get physics
off the ground.  But in the end they were forms of dogma that had to
be discarded.

The beautiful thing about physics is that it has an orderly way to
conduct revolutions, and to discard older ideas; or more accurately,
to assign older ideas to their proper domain of limited significance,
while moving on to more accurate descriptions of reality.  I expect
that a thorough exploration of the presence and ways of functioning of
the subject, on a par with that of objects, will revolutionize science
in a constructive way, within this century and the next.

\section{Conclusion}
\label{sec:concl}

We have discussed the nature of reality in the ontological context of
Penrose's math-matter-mind triangle.  Physicists have widely differing
views of the 
deeper meaning of physics,
and we have found that we three authors reflect this disparity.
Despite
our similarities in research interests, knowledge and cultural
background, we espouse conflicting views that we have termed
fundamentalist, secular, and mystic.  A key message for non-physicists
reading this paper is therefore that they should 
be deeply suspicious of any
self-proclaimed popularizer or other ambassador claiming to speak on
these matters on behalf of the consensus of the theoretical physics
community. 

A second noteworthy conclusion is that despite these disagreements, 
we advocate rather similar paths forward.
This illustrates that physicists can vary strongly in their 
interpretations of the meaning of physics, while agreeing quite well on how 
it should be done.

Looking to the immediate future of physics,
one thing that we agree on is that the diversity in views in the
physics community is healthier than an ontological monoculture.
Our best hope of making progress on  the 
open questions of physics is to tackle them with
a wide variety of ideas.

\smallskip
\noindent {\bf Acknowledgments}\\
The authors are grateful to the Institute for Advanced Study in
Princeton for providing the stimulating environment for our early
discussions, which over the past eight years evolved into this paper.

\end{document}